\documentclass[%
 reprint,
 amsmath,amssymb,
 aps,
]{revtex4-2}

\usepackage{graphicx}
\usepackage{dcolumn}
\usepackage{bm}
\usepackage{orcidlink}

\begin{document}

\preprint{APS/123-QED}
\newcommand{\ket}[1]{\left\vert{#1}\right\rangle}
\newcommand{\bra}[1]{\left\langle{#1}\right\vert}
\newcommand{\abs}[1]{\left\vert{#1}\right\vert}
\title{Tunable photon-photon correlations in waveguide QED systems with giant atoms}

\author{Wenju Gu~\orcidlink{0000-0002-6173-6885}}
\email{guwenju@yangtzeu.edu.cn}
\altaffiliation{School of Physics and Optoelectronic Engineering, Yangtze University, Jingzhou 434023, China}
\author{Lei Chen}%
\affiliation{School of Physics and Optoelectronic Engineering, Yangtze University, Jingzhou 434023, China}%
\author{Zhen Yi}%
\affiliation{School of Physics and Optoelectronic Engineering, Yangtze University, Jingzhou 434023, China}
\author{Sujing Liu}%
\affiliation{School of Physics and Optoelectronic Engineering, Yangtze University, Jingzhou 434023, China}

\author{Gao-xiang Li}
\email{gaox@mail.ccnu.edu.cn}
\affiliation{Department of Physics, Huazhong Normal University, Wuhan 430079, China}%

\begin{abstract}
We investigate the scattering processes of two photons in a one-dimensional waveguide coupled to two giant atoms. By adjusting the accumulated phase shifts between the coupling points, we are able to effectively manipulate the characteristics of these scattering photons. Utilizing the Lippmann-Schwinger formalism, we derive analytical expressions for the wave functions describing two-photon interaction in separate, braided, and nested configurations. Based on these wave functions, we also obtain analytical expressions for the incoherent power spectra and second-order correlation functions. In contrast to small atoms, the incoherent spectrum, which is defined by the correlation of the bound state, can exhibit more tunability due to the phase shifts. Additionally, the second-order correlation functions in the transmission and reflection fields could be tuned to exhibit either bunching or antibunching upon resonant driving. These unique features offered by the giant atoms in waveguide QED could benefit the generation of nonclassical itinerant photons in quantum networks.
\end{abstract}

\maketitle

\section{Introduction}
Waveguide quantum electrodynamics (QED) have garnered significant interest due to the emergence of unique physical phenomena when atom-photon coupling to a continuum of modes limited to a single dimension~\cite{RevModPhys.87.347,PhysRevA.94.043839}, as well as for their applications in quantum networks~\cite{Kimble2008,PhysRevA.74.043818,sciadv.abb8780,science.1155441}. With advancements in technology, high coupling efficiency between atomic degrees of freedom (including natural and artificial atoms) and propagating photonic modes has been realized within different state-of-the-art platforms~\cite{Mirhosseini2019,Kannan2023,RevModPhys.95.015002}. Then, energy dissipation from the atom into the waveguide dominates over that into modes other than the waveguide~\cite{RevModPhys.95.015002}. By entering the high coupling efficiency regime, atoms can function as high-quality quantum emitters, enabling demonstration of primitives of quantum networks~\cite{PhysRevLett.107.073601,Hoi_2013}. Furthermore, there has been a shift towards investigating multi-atom phenomena in waveguide QED, such as correlated dissipation, waveguide-mediated interactions between multiple atoms, and many-body phenomena~\cite{PhysRevA.88.043806,science.1244324,PhysRevApplied.15.044041,PhysRevA.104.053703}. Another interesting effect in waveguide QED is related to photonic modes. The optical nonlinearity becomes apparent on the scale of a few photons, allowing for observation of quantum nonlinear phenomena through optical correlation functions~\cite{PhysRevA.87.063818,PhysRevA.93.033856,PhysRevB.81.155117,PhysRevLett.106.053601,PhysRevA.106.043722, andp.202200512}. One manifestation of the nonlinearity is the presence of two- and higher-order photon bound states~\cite{PhysRevLett.98.153003,RevModPhys.95.015002}. In these bound states, photons strongly exhibit correlations, meaning that once one photon is detected, the arrival of another photon is much more likely compared to a random time. It is important to note that photon bound states are distinct from bunched photon states. Photon bound states are quasiparticles with their own dispersion and are eigenstates of the underlying Hamiltonian that describes the nonlinear medium~\cite{Tomm2023}.

In recent years, a new paradigm in quantum optics has emerged, beyond the dipole approximation in the light-atom interaction. This paradigm challenges the assumption that the size of atoms is significantly smaller than the wavelength of the interaction light, giving rise to the concept of ``giant atoms''. Giant atoms can couple to light or other bosonic fields at multiple points, which may be spaced wavelengths apart. Such systems can be implemented both with superconducting qubits coupled either to microwave transmission lines~\cite{Kannan2020} or surface acoustic waves~\cite{science.1257219}. The study of giant atom can be divided into two categories: Markovian and non-Markovian regimes. In the Markovian case, the propagation time for radiation across the atom is much shorter than the interaction time with the atom. Multiple coupling points in giant atoms give rise to interference effects, allowing for a coherent exchange interaction between atoms mediated by a waveguide. This can result in effects such as frequency-dependent couplings, Lamb shifts, and relaxation rates~\cite{PhysRevA.103.023710,PhysRevA.90.013837}.  On the other hand, giant atoms in the non-Markovian regime interact with the radiation field at a timescale comparable to that for radiation to propagate across the atom, resulting in effects such as nonexponential decay~\cite{Andersson2019, S.Longhi@ol2020, S.Guo@pra2020,Qiu2023} and oscillating bound states~\cite{PhysRevResearch.2.043014}. Extending the concept of multiple small atoms to multiple giant atoms enables the exploration of a diverse and rich range of phenomena. These include waveguide-mediated decoherence-free subspaces~\cite{PhysRevLett.120.140404,PhysRevResearch.2.043184} and the enhanced spontaneous sudden birth of entanglement~\cite{PhysRevLett.130.053601}. The simplest configuration for studying these phenomena involves two giant atoms interacting to a waveguide with two coupling points. These layouts can be categorized into three distinct configurations based on the arrangement of the coupling points: separate, braided, and nested~\cite{PhysRevLett.120.140404}. While most studies have focused on the atomic degrees of freedom~\cite{PhysRevA.106.063703,PhysRevA.108.023728,PhysRevLett.130.053601,PhysRevLett.120.140404,PhysRevResearch.2.043184}, there have been some investigations into the photonic degrees of freedom. However, these studies primarily explore the single-excitation subspace~\cite{PhysRevA.104.063712,PhysRevA.108.043709,PhysRevA.107.063703}. To the best of our knowledge, the optical nonlinearity involving two or multiple photons in giant atoms is less investigated.

The phenomenon of a single two-level atom not being able to emit two photons simultaneously is widely known. This limitation arises due to the fact that the atom can absorb only one photon at a time, resulting in $g^{(2)}(0)=0$ in the reflection channel~\cite{PhysRevA.76.062709,Chang2007}. By introducing multiple atoms, the constraint can be overcome, and it becomes possible to manipulate the correlation between photons~\cite{PhysRevA.91.053845}. This occurs because when one photon becomes trapped within the first atom, there is a probability that the second photon will propagate to and reflect off the subsequent atom. This process leads to the simulated emission of the first photon, effectively allowing the simultaneous emission of two photons. Therefore, the probability of two photons being emitted together is not completely prohibited. A similar scenario unfolds with two giant atoms, exhibiting even more pronounced effects. In this work, we employ the Lippmann-Schwinger (LS) formalism~\cite{H.Zheng@prl2013,Y.Fang@EPJ2014,PhysRevA.83.043823} to analyze the two-photon scattering processes involving two giant atoms coupled to a one-dimensional (1D) waveguide, which contains separate, braided, and nested configurations. By utilizing this approach, we are able to obtain the analytical two-photon interacting scattering wavefunctions for three configurations. Additionally, the incoherent power spectrum is derived from the correlation of the bound state, with its total flux serving as an indicator of photon-photon correlation. The second-order correlation function provides a direct measure of photon-photon correlation. Through our analysis, we find that the accumulated phase shifts can be utilized to manipulate the photon-photon correlation and the evolution of the second-order correlation for photons scattered by the giant atoms.

The paper is organized as follows. In Sec.~\ref{sec2},  we present the physical model that describes the waveguide QED system with two giant atoms. Sections~\ref{sec3} and~\ref{sec4} derive the single photon scattering eigenstates and two-photon interacting eigenstates in the three configurations. Sections~\ref{sec5} and \ref{sec6} analyze the incoherent power spectra and the second-order correlation functions. The conclusions drawn from our study are given in Sec.~\ref{conclusion}.

\section{Physical model}
\label{sec2}
\begin{figure}[hbt]
\centering\includegraphics[width=7cm,keepaspectratio,clip]{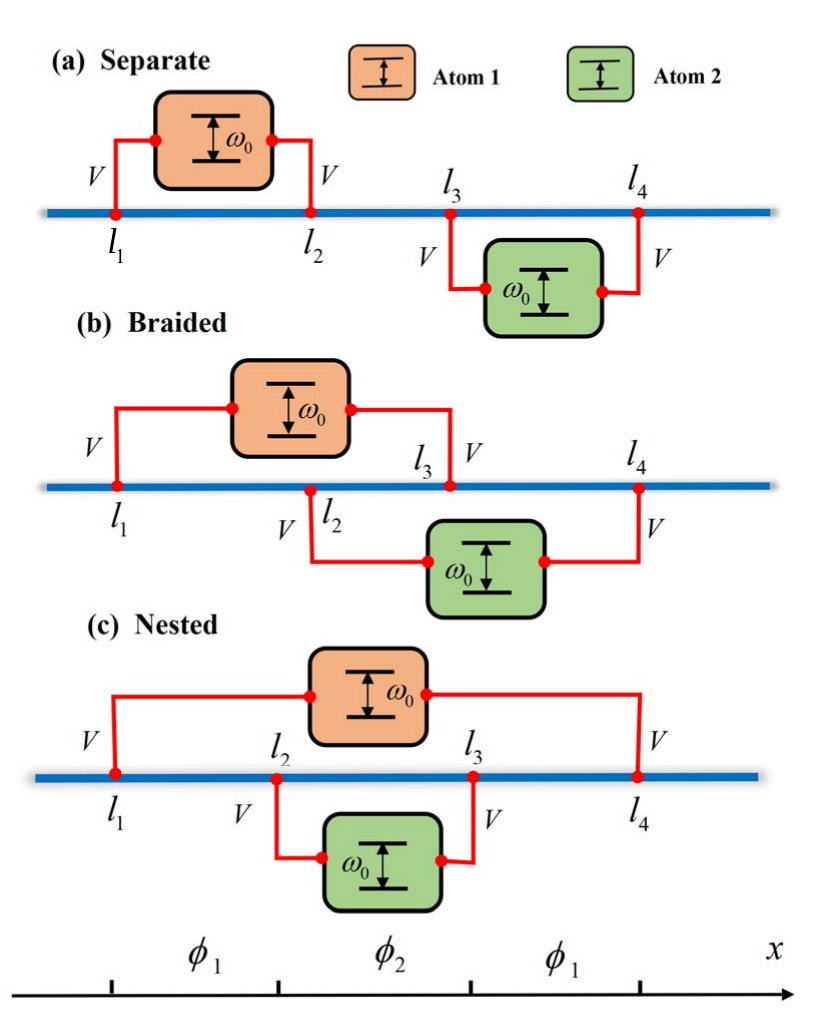}
\caption{Schematic illustration of two two-level giant atoms with a one-dimensional (1D) waveguide in three distinct configurations: (a) separate, (b) braided, (c) nested. The coupling between the atomic transitions and the waveguide modes occurs at four specific points denoted as $l_k$, where $k=1,2,3,4$. The strength of the coupling is represented by $V$. Additionally, the phase shifts acquired between neighboring points are represented by $\phi_1$ and $\phi_2$.}
\label{fig1}
\end{figure}
We consider a waveguide QED system composed of two two-level giant atoms coupled to an open 1D waveguide. Each giant atom only interacts with the waveguide at two coupling points, allowing for three distinct configurations: separate, braided, and nested, as illustrated in Fig.~\ref{fig1}. The Hamiltonian describing the system in real space is given by ($\hbar=1$ hereafter):
\begin{align}
\hat{H}^{c}=&\left(\omega_0-i\frac{\gamma_e}{2}\right)\sum_{j=1}^2\hat{\sigma}^{+}_j\hat{\sigma}^{-}_j\nonumber\\&-iv_g\int dx\left[\hat{a}_R^\dagger(x)\partial_x\hat{a}_R(x)-\hat{a}_L^\dagger(x)\partial_x\hat{a}_L(x)\right]\nonumber\\&
+\sum_{\alpha=R,L}\sum_{j=1}^2V\int dxQ_j^{c}(x)\left[\hat{a}_\alpha^\dagger(x)\hat{\sigma}_j^{-}+\hat{a}_\alpha(x)\hat{\sigma}_j^+\right],  
\end{align}
where $c=s,b,n$ denotes the separate, braided and nested configurations, respectively. Here, the two giant atoms are assumed to be identical with the same transition frequency $\omega_0$ and dissipation rate $\gamma_e$. The excited and ground states of the $j$th giant atom are represented by $\ket{e}_j$ and $\ket{g}_j$, respectively. The atomic raising and lowering operators are denoted as $\hat{\sigma}_j^+=\ket{e}_{jj}\bra{g}$ and $\hat{\sigma}_j^-=\ket{g}_{jj}\bra{e}$, respectively. Additionally, $\hat{a}_R(x)$ and $\hat{a}_L(x)$ correspond to the annihilation operators of right-moving and left-moving photons in the waveguide, and $\nu_g$ is the group velocity. For simplicity, we set $\nu_g=1$ in the following. The coupling between the giant atoms and the waveguide occurs at connection points identified by $Q_j^{c}(x)$, with a common coupling strength $V$. In the case of separate configuration, $Q_1^{s}(x)=\delta(x-l_1)+\delta(x-l_2)$ refers to the coupling points $l_1$ and $l_2$ of the first giant atom, while $Q_2^{s}(x)=\delta(x-l_3)+\delta(x-l_4)$ refers to the coupling points $l_3$ and $l_4$ of the second giant atom. Similarly, for the braided configuration, $Q_1^{b}(x)=\delta(x-l_1)+\delta(x-l_3)$, and $Q_2^{b}(x)=\delta(x-l_2)+\delta(x-l_4)$. In the nested configuration, $Q_1^{n}(x)=\delta(x-l_1)+\delta(x-l_4)$ and $Q_2^{n}(x)=\delta(x-l_2)+\delta(x-l_3)$. To exploit the benefits of parity symmetry, the positions of the atoms can be deliberately selected to exhibit symmetry with respect to the origin, i.e., $l_1=-l_4$ and $l_2=-l_3$. The phase shifts acquired between neighboring coupling points are given by $\phi_1=k(l_2-l_1)=k(l_4-l_3)$ and $\phi_2=k(l_3-l_2)$.

The total excitation number of the system is conserved in the interaction, and thus in the single-excitation subspace, the eigenstate can be written in the form
\begin{align}
\vert\Phi_1^{c}(k)\rangle_\alpha=&\Bigg\{\int dx\left[{\phi_R^\alpha}(k,x)\hat{a}^\dagger_R(x)+{\phi_L^\alpha}(k,x)\hat{a}_L(x)\right]\nonumber\\&+\sum_{j=1}^2e_{j\alpha}^{c}(k)\hat{\sigma}_j^\dagger\Bigg\}\vert0\rangle,
\end{align}
where $\alpha$ refers to the direction of incoming photons, and $\phi_{R/L}^{\alpha}(k,x)$ denote the probability amplitudes of creating the right-moving and left-moving photons in real space for the $\alpha$-direction incident photon with wave vector $k$, respectively. Furthermore, $e_{j\alpha}^{c}(k)$ is the excitation amplitude of the $j$th atom in the $c$ configuration, and $\ket{0}$ represents the vacuum state of the system. The probability amplitudes are determined by the Schr\"{o}dinger equation $\hat{H}^{c}\vert\Phi_1^{c}(k)\rangle_\alpha=k\vert\Phi_1^{c}(k)\rangle_\alpha$, which fulfill
\begin{align}
&\left(\omega_0-i\frac{\gamma_e}{2}-k\right)e_{1\alpha}^c(k)+V\sum_{\alpha^\prime}\int dx Q_1^{c}(x)\phi_{\alpha^\prime}^\alpha(k,x)=0,\nonumber\\
&\left(\omega_0-i\frac{\gamma_e}{2}-k\right)e_{2\alpha}^c(k)+V\sum_{\alpha^\prime}\int dxQ_2^{c}(x)\phi_{\alpha^\prime}^\alpha(k,x)=0,\nonumber\\
&\left(-i\partial_x-k\right)\phi_R^\alpha(k,x)+V\sum_{j=1}^2Q_j^{c}(x)e_{j\alpha}^c(k)=0,\nonumber\\
&\left(i\partial_x-k\right)\phi_L^\alpha(k,x)+V\sum_{j=1}^2Q_j^{c}(x)e_{j\alpha}^c(k)=0.
\label{eqn:Schrodinger1}
\end{align}
The solutions of three different configurations will be presented explicitly in the following.

\section{Single-photon scattering eigenstates}
\label{sec3}
In this section, we present the eigenstates of single-photon scattering for each of the three coupling configurations. These eigenstates contain the amplitudes of atomic excitation, as well as the amplitudes for single-photon transmission and reflection.

\subsection{Separate-coupling case}
In the separate configuration depicted in Fig.~\ref{fig1}(a), when a photon is injected in the right-moving direction (i.e., $\alpha=R$) with wave vector $k$, the amplitudes can be concretely expressed in the form
\begin{align}
\phi_R^R(k,x)=&\frac{e^{ikx}}{\sqrt{2\pi}}\Bigg[\theta(l_1-x)+\sum_{i=1}^3t_i^s(k)\theta(x-l_i)\theta(l_{i+1}-x)\nonumber\\&+t_4^s(k)\theta(x-l_4)\Bigg],\nonumber\\
\phi_L^R(k,x)=&\frac{e^{-ikx}}{\sqrt{2\pi}}\Bigg[r_1^s(k)\theta(l_1-x)\nonumber\\&+\sum_{i=2}^4r_i^s(k)\theta(x-l_{i-1})\theta(l_i-x)\Bigg].
\end{align}
Within the symmetric topology, by substituting these coefficients into Eqs.~\eqref{eqn:Schrodinger1}, we can derive the solutions for transmission and reflection amplitudes as well as atomic excitation amplitudes as follows:
\begin{align}
t_4^s(k)=&(k-\omega_0-\Gamma\sin\phi_1)^2/D^s,\nonumber\\
r_1^s(k)=&-4i\Gamma\cos^2\frac{\phi_1}{2}\big\{(k-\omega_0)\cos(\phi_1+\phi_2)\nonumber\\&+\Gamma[\sin\phi_2+\sin(\phi_1+\phi_2)]\big\}/D^s,\nonumber\\
e_{1R}^s(k)=&\frac{e^{-i\phi_2/2}}{2}\sqrt{\frac{\Gamma}{\pi}}(1+e^{-i\phi_1})\Big\{k-\omega_0+i\Gamma(1+e^{i\phi_1})\nonumber\\&
-i\frac{\tilde{\Gamma}_s}{2}e^{i(\phi_1+\phi_2)}\Big\}/D^s,\nonumber\\
e_{2R}^s(k)=&\frac{e^{i\phi_2/2}}{2}\sqrt{\frac{\Gamma}{\pi}}(1+e^{i\phi_1})(k-\omega_0-\Gamma\sin\phi_1)/D^s,\nonumber\\
D^s=&\left[\omega_0-k-i\Gamma(1+e^{i\phi_1})\right]^2+\frac{{{\tilde{\Gamma}_s}^2}}{4}.
\end{align}
Here $\Gamma=2V^2$ represents the decay rate of atomic dissipation to the waveguide continuum, and $\tilde{\Gamma}_s=\Gamma e^{i\phi_2}(1+e^{i\phi_1})^2$. In the high coupling efficiency regime~\cite{RevModPhys.89.021001}, the spontaneous decay rate to the waveguide dominates over the decay to other modes, i.e., $\Gamma\gg\gamma_e$. Consequently, $\gamma_e$ can been ignored in the following discussions.

According to the parity symmetry, for a photon injected in the left-moving direction (i.e., $\alpha=L$), the transmission and reflection amplitudes are equivalent to those of the right-moving case. In addition, the atomic excitation amplitudes also follow this symmetry, which fulfill
\begin{align}
 e_{2L}^s(k)=e_{1R}^s(k), \hspace{5pt} e_{1L}^s(k)=e_{2R}^s(k).
\end{align}

\subsection{Braided-coupling case}
Next, let us consider the braided-coupling case, as shown in Fig.~\ref{fig1}(b). In this configuration, the coupling points are denoted by $Q_j^{b}(x)$. Following the same procedure employed in the separate-coupling case, one can determine the corresponding transmission and reflection amplitudes, as well as the atomic excitation amplitudes, which are
\begin{align}
t_4^b(k)=&\big[(k-\omega_0)^2-2\Gamma(k-\omega_0)\sin(\phi_1+\phi_2)\nonumber\\&+\Gamma^2\sin\phi_1(\sin\phi_1-2\sin\phi_2)\big]/D^b,\nonumber\\
r_1^b(k)=&-4i\Gamma\cos^2\frac{\phi_1+\phi_2}{2}\big[(k-\omega_0)\cos\phi_1\nonumber\\&+\Gamma\sin\phi_1\big]/D^b,\nonumber\\
e_{1R}^b(k)=&\frac{e^{i\phi_2/2}}{2}\sqrt{\frac{\Gamma}{\pi}}\left[1+e^{-i(\phi_1+\phi_2)}\right]\Big[k-\omega_0\nonumber\\&-i\frac{\Gamma}{2}\left(-1+e^{i2\phi_1}\right)\left(2+e^{i(\phi_1+\phi_2)}\right)\Big]/D^b,\nonumber\\
e_{2R}^b(k)=&\frac{e^{i\phi_2/2}}{2}\sqrt{\frac{\Gamma}{\pi}}\left[1+e^{-i(\phi_1+\phi_2)}\right]\Big[e^{i\phi_1}(k-\omega_0)\nonumber\\&+i\frac{\Gamma}{2}e^{i\phi_2}\left(-1+e^{i2\phi_1}\right)\Big]/D^b,\nonumber\\    
D^b=&\left[\omega_0-k-i\Gamma\left(1+e^{i(\phi_1+\phi_2)}\right)\right]^2+\frac{\tilde{\Gamma}_b^2}{4}.
\end{align}
where $\tilde{\Gamma}_b=\Gamma\left[2e^{i\phi_1}+e^{i\phi_2}+e^{i(2\phi_1+\phi_2)}\right]$. Also owing to the presence of parity symmetry,  for the case of left-moving photon injection, the transmission and reflection amplitudes remain equivalent to those in the right-moving scenario. The atomic excitation amplitudes manifest as $e_{2L}^b(k)=e_{1R}^b(k)$ and $e_{1L}^b(k)=e_{2R}^b(k)$.
\subsection{Nested-coupling case}
Finally, we turn to the nested-coupling case, as shown in Fig.~\ref{fig1}(c). In this configuration, the corresponding coupling points are denoted by $Q_j^{n}(x)$. Employing the same procedure, we can derive the transmission and reflection amplitudes, as well as atomic excitation amplitudes as
\begin{align}
t_4^n(k)=&\Big\{\left(k-\omega_0-\Gamma\sin\phi_2\right)\left[k-\omega_0-\Gamma\sin(2\phi_1+\phi_2)\right]\nonumber\\&-\Gamma^2\left[\sin\phi_1+\sin(\phi_1+\phi_2)\right]^2\Big\}/D^n,\nonumber\\
r_1^n(k)=&-2i\Gamma\Big\{(k-\omega_0)\left[1+\cos\phi_1\cos(\phi_1+\phi_2)\right]\nonumber\\&+\Gamma\sin\phi_1\left[\cos\phi_1+\cos(\phi_1+\phi_2)\right]\Big\}/D^n,\nonumber\\
e_{1R}^n(k)=&\sqrt{\frac{\Gamma}{\pi}}\Big[(k-\omega_0)\cos\left(\phi_1+\frac{\phi_2}{2}\right)\nonumber\\&+2\Gamma\sin\phi_1\cos\frac{\phi_2}{2}\Big]/D^n,\nonumber\\
e_{2R}^n(k)=&\sqrt{\frac{\Gamma}{\pi}}(k-\omega_0)\cos\frac{\phi_2}{2}/D^n,\nonumber\\
D^n=&\left[\omega_0-k-i\frac{\Gamma}{2}\left(2+e^{i\phi_2}+e^{i(2\phi_1+\phi_2)}\right)\right]^2+\frac{\tilde{\Gamma}_n^2}{4}.
\end{align}
where $\tilde{\Gamma}_n=\Gamma\sqrt{e^{i2\phi_2}(1+e^{i2\phi_1})^2+4e^{i2\phi_1}(1+2e^{i\phi_2})}$. In the presence of parity symmetry, for the left-moving injection of a photon, i.e., $\alpha=L$, the transmission and reflection amplitudes are the equivalent to those of right-moving case. Additionally, the atomic excitation amplitudes satisfy $e_{1L}^n(k)=e_{1R}^n(k)$ and $e_{2L}^n(k)=e_{2R}^n(k)$, which differ from those obtained in the separate-coupling and braided-coupling cases. This is because in the nested configuration, the atoms remain unchanged for the left-moving incident photon, whereas they are exchanged in the separate and braided configurations. Concretely, by defining a parity operator $\hat{P}$, $\hat{P}\hat{\sigma}_j\hat{P}^\dagger=\hat{\sigma}_{3-j}$ for the separate and braided cases, while $\hat{P}\hat{\sigma}_j\hat{P}^\dagger=\hat{\sigma}_{j}$ for the nested case~\cite{Dinc2019exactmarkoviannon}.

\section{Two-photon interacting scattering eigenstates}
\label{sec4}
By utilizing the obtained eigenstates for the single-photon excitation, we can proceed to construct the two-photon interacting eigenstates via employing LS techniques~\cite{PhysRevLett.110.113601,Y.Fang@EPJ2014,PhysRevA.83.043823,PhysRevA.108.053718}. The construction is given by
\begin{align}
\vert\Psi_2^{c}(k_1,k_2)\rangle_{\alpha_1\alpha_2}=&\vert\Phi_2^{c}(k_1,k_2)\rangle_{\alpha_1\alpha_2}\nonumber\\
&+\hat{G}^R(E)\hat{V}_\text{on}\vert\Phi_2^{c}(k_1,k_2)_{\alpha_1\alpha_2}.
\end{align}
Here, $\hat{G}^R(E)$ represents the retarded Green's function, and $E=k_1+k_2$ corresponds to the total energy of two incident photons. Moreover, $\hat{V}_\text{on}$ denotes the on-site interaction in the bosonic representation of the atoms. In real space, this construction can be expressed as
\begin{align}
&_{\alpha_1^\prime\alpha_2^\prime}\bra{x_1x_2}\Psi_2^{c}(k_1,k_2)\rangle_{\alpha_1\alpha_2}=_{\alpha_1^\prime\alpha_1^\prime}\bra{x_1x_2}\Phi_2^{c}(k_1,k_2)\rangle_{\alpha_1\alpha_2}\nonumber\\
&-\sum_{i,j=1}^2G_{i,c}^{\alpha_1^\prime\alpha_2^\prime}(x_1,x_2)(G_c^{-1})_{ij}\bra{d_jd_j}\Phi_2^{c}(k_1,k_2)_{\alpha_1\alpha_2},
\label{eqn:LS}
\end{align}
where $\ket{\Phi_2^{c}(k_1,k_2)}_{\alpha_1\alpha_2}=\frac{1}{\sqrt{2}}\ket{\Phi_1^{c}(k_1)}_{\alpha_1}\otimes\ket{\Phi_1^{c}(k_2)}_{\alpha_2}$ is the two-photon non-interacting eigenstate. In order to obtain the interacting eigenstates, it is crucial to derive the elements of the Green's function, which are provided as follows:
\begin{align}
&G_{i,c}^{\alpha_1\alpha_2}(x_1,x_2)=\sum_{\alpha_1^\prime\alpha_2^\prime}\int dk_1dk_2\nonumber\\&\times\frac{_{\alpha_1\alpha_2}\bra{x_1x_2}\Phi_2^c(k_1,k_2)\rangle_{\alpha_1^\prime\alpha_2^\prime}\bra{\Phi_2^c(k_1,k_2)}d_id_i\rangle}{E-k_1-k_2+i0^+},\nonumber\\
&G_{ij}^c=\sum_{\alpha_1\alpha_2}\int dk_1dk_2\nonumber\\&\times\frac{\langle d_id_i\vert\Phi_2^c(k_1,k_2)\rangle_{\alpha_1,\alpha_2}\langle\Phi_2^c(k_1,k_2)\vert d_jd_j\rangle}{E-k_1-k_2+i0^+},\nonumber\\
&G^{-1}_c=\left(
    \begin{array}{cc}
       G_{11}^c  & G_{12}^c \\
       G_{21}^c  & G_{22}^c
    \end{array}
    \right)^{-1}.
\end{align}
It should be noted that $x_1$ and $x_2$ refer to the positions of the photons. Upon examining the structure of these Green's functions, one can find the presence of two distinct components which are
\begin{align}
_{\alpha_1^\prime\alpha_2^\prime}\langle x_1x_2\ket{\Phi_2^c(k_1,k_2)}_{\alpha_1\alpha_2}=&\frac{1}{2}\Big[\phi_{\alpha_1^\prime}^{\alpha_1}(k_1,x_1)\phi_{\alpha_2^\prime}^{\alpha_2}(k_2,x_2)\nonumber\\&+\phi_{\alpha_1^\prime}^{\alpha_2}(k_2,x_1)\phi_{\alpha_2^\prime}^{\alpha_1}(k_1,x_2)\Big], \nonumber\\
\langle d_id_i\ket{\Phi_2^c(k_1,k_2)}_{\alpha_1\alpha_2}=&e_{i{\alpha_1}}^c(k_1)e_{i{\alpha_2}}^c(k_2).
\end{align}
In principle, the accumulated phase shifts between the coupling points are dependent on the wave vector, which introduces significant complexity into the photon scattering processes~\cite{PhysRevLett.110.113601,PhysRevA.95.053821,Dinc2019exactmarkoviannon,PhysRevA.108.053718}. However, for the purposes of this study, we focus on the Markovian approximation $\tau\Gamma\ll1$~\cite{PhysRevA.95.053821}, where $\tau$ is the propagation time across the giant atoms. Consequently, we explicitly substitute the frequency with the atomic transition frequency $\omega_0$, resulting in phase factors denoted as $\phi_1=k_0(l_2-l_1)=k_0(l_4-l_3)$ and $\phi_2=k_0(l_3-l_2)$ where $k_0=\omega_0/\nu_g$. Finally, under the assumption of $x_1>l_4$ and $x_2=x_1+x$ (with $x>0$), the two-photon interacting eigenstate of the injection in right-moving direction can be expressed as
\begin{align}
\ket{\Psi_2^c(k_1,k_2)}_{RR}=&\int dx_1dx_2\Bigg[\frac{f_{RR}^c(x_1,x_2)}{\sqrt{2}}\hat{a}^\dagger_R(x_1)\hat{a}^\dagger_R(x_2)\nonumber\\
&+f_{RL}^c(x_1,x_2)\hat{a}^\dagger_R(x_1)\hat{a}^\dagger_L(x_2)\nonumber\\
&+\frac{f_{LL}^c(x_1,x_2)}{\sqrt{2}}\hat{a}^\dagger_L(x_1)\hat{a}^\dagger_L(x_2)\Bigg]\ket{0}.
\end{align}
The coefficients $f_{\alpha_1\alpha_2}^c(x_1,x_2)$ can be written in a common form, which consists of a two-particle plane wave with rearranged momenta of the photons and a bound state. The emergence of the plane wave is attributed to coherent scattering, while the bound state arises from incoherent scattering. It is worth noting that the bound state exhibits exponential decay as the distance between the two photons increases. This phenomenon is closely related to the two-particle irreducible $T$-matrix in scattering theory~\cite{PhysRevLett.111.223602}.
Therefore, the two-photon transmission and reflection amplitudes can be written in the form
\begin{align}
f_{RR}^c(x_1,x_2)=&\frac{e^{iEx_c}}{\sqrt{2}\pi}\left[t_4^c(k_1)t_4^c(k_2)\cos\Delta_1x+B_{k_1k_2}^{RR,c}(x)\right],\nonumber\\
f_{LL}^c(x_1,x_2)=&\frac{e^{-iEx_c}}{\sqrt{2}\pi}\left[r_1^c(k_1)r_1^c(k_2)\cos\Delta_1x+B_{k_1k_2}^{LL,c}(x)\right],\nonumber\\
\label{eqn:two-photon_amp}
\end{align}
where $x_c=(x_1+x_2)/2$ is the center position of the two photons, and $\Delta_1=(k_1-k_2)/2$ represents half of the energy difference between the two incident photons. The explicit expressions of the bound-state terms in three configurations are given in Appendix A.

\section{Incoherent power spectrum}
\label{sec5}
The two-photon interacting eigenstate comprises two components, namely the plane wave resulting from coherent scattering and the bound state arising from photon-photon interactions. In order to examine the effect of the bound state on scattering processes, our initial focus is directed towards the power spectrum or resonance fluorescence, which can be obtained by performing a Fourier transform of the first-order correlation function,
\begin{align}
S^{\alpha,c}(\omega)=\int dt e^{-i\omega t}\bra{\Psi_2^c}\hat{a}^\dagger_\alpha(x_0)\hat{a}_\alpha(x_0+t)\ket{\Psi_2^c},
\label{eqn:power_spectrum}
\end{align}
where $x_0$ represents the position of a distant detector located outside the scattering region. $S^{\alpha,c}(\omega)$ accounts for the spectral decomposition of the photons in the interacting two-photon wave function $\ket{\Psi_2^c}$. In general, the power spectrum consists of the coherent and incoherent parts, i.e., $S^{\alpha,c}(\omega)=S^{\alpha,c}_{\text{coh}}(\omega)+S^{\alpha,c}_{\text{incoh}}(\omega)$. The contribution from coherent scattering manifests as a $\delta$ function, while the correlation of the bound state within the wave function accounts for the incoherent scattering,
\begin{align}
S^{\alpha,c}_\text{incoh}(\omega)=&\frac{1}{\pi^2}\int dtdxe^{i(E/2-\omega)t}{B_{k_1k_2}^{\alpha\alpha,c*}}(x)B_{k_1k_2}^{\alpha\alpha,c}(x-t).
\label{eqn:sincoh}
\end{align}

Via substituting the expressions of the bound state, the incoherent power spectra in the transmission and reflection can be written in the form
\begin{align}
S^{R,c}_\text{incoh}(\omega)=\frac{1}{\pi^2}\abs{M_R^c(\omega)}^2,\hspace{5pt}
S^{L,c}_\text{incoh}(\omega)=\frac{1}{\pi^2}\abs{M_L^c(\omega)}^2,
\end{align}
where
\begin{align}
M_R^c(\omega)=&Z_1^{c}A_1^{c}(E/2-\omega)+Z_2^cA_2^c(E/2-\omega),\nonumber\\
M_L^c(\omega)=&Z_3^{c}A_1^{c}(E/2-\omega)+Z_4^cA_2^c(E/2-\omega),
\end{align}
and
\begin{align}
A_1^c(y)=&\frac{1}{iy-i\eta_c/2+\tilde{\Gamma}_c/2}+(y\leftrightarrow-y),\nonumber\\
A_2^c(y)=&\frac{1}{iy-i\eta_c/2-\tilde{\Gamma}_c/2}+(y\leftrightarrow-y).
\end{align}
To explore physical implications of the incoherent power spectra, we perform a Fourier transform of the transmitted and reflected states from their real space representation to frequency space. In frequency space, these states can be expressed as $t_4^c(\omega_1)t_4^c(\omega_2)\hat{a}_R(\omega_1)\hat{a}_R(\omega_2)\ket{0}+\frac{1}{2\pi}\int d\omega M_R^c(\omega)\hat{a}_R(E-\omega)\hat{a}_R(\omega)\ket{0}$ in transmission and $r_1^c(\omega_1)r_1^c(\omega_2)\hat{a}_L(\omega_1)\hat{a}_L(\omega_2)\ket{0}+\frac{1}{2\pi}\int d\omega M_L^c(\omega)\hat{a}_L(E-\omega)\hat{a}_L(\omega)\ket{0}$ in reflection. Within each of these expressions, the first term describes the independent propagation of the two photons, while the second term represents the formation of the bound state between the two photons after undergoing inelastic scattering. According to the principle of energy conservation, the scattered photons are always generated in pairs with frequencies of opposite signs. The coefficients $M_R^c(\omega)$ and $M_L^c(\omega)$ serve to quantify the production of these photon pairs in the transmission and reflection processes~\cite{PhysRevLett.84.5304}. Therefore, the incoherent power spectrum can provide a direct measure of the generation of photon pairs at the frequency $\omega$.

\begin{figure}[hbt]
\centering\includegraphics[width=8cm,keepaspectratio,clip]{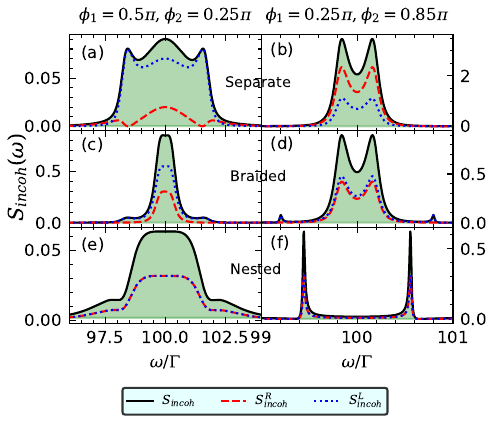}
\caption{Incoherent power spectra in the three different configurations as a function of frequency $\omega$ with different values of $\phi_1$ and $\phi_2$. The first row corresponds to the separate case, the second row corresponds to the braided case, and the third row corresponds to the nested case. In addition, the black solid lines denote the total incoherent power spectra, the red dashed lines denote the incoherent power spectra in transmission, and the blue dotted lines denote the incoherent power spectra in reflection. The other parameters are $k=\omega_0=100\Gamma$.}
\label{fig2}
\end{figure}

Under the assumption of a narrow bandwidth of incident photons, where the spectral width of the wave packet is significantly smaller than $\Gamma$, the wave packet can be approximated as a $\delta$ function. This implies that the incident photons have an equal frequency $k_1=k_2=k=E/2$. In this case, the incoherent power spectra, including transmission $S^{R,c}_\text{incoh}(\omega)$, reflection $S^{L,c}_\text{incoh}(\omega)$, and total spectrum $S^{c}_\text{incoh}(\omega)=\sum_\alpha S^{\alpha,c}_\text{incoh}(\omega)$ (transmission + reflection), are plotted in Fig.~\ref{fig2} as a function of $\omega$. In these figures, the phase shifts between coupling points can be engineered through device design, particularly by adjusting the relative lengths of the waveguide segment. The selected values of $\phi_1=0.5\pi$ and $\phi_2=0.25\pi$ are based on the experimental setup described in Ref.~\cite{Kannan2020}. To cover different scenarios, we also consider an alternative case with $\phi_1=0.25\pi$ and $\phi_2=0.85\pi$. These choice of phase shifts allow us to explore various scenarios and investigate their impact on the incoherent power spectra.

It can be verified that the location and the width of the peak are determined by the real and imaginary parts of the roots of the denominators in $A_1^c(E/2-\omega)$ and $A_2^c(E/2-\omega)$. For the two giant atoms, these properties can be adjusted through the accumulated phase shifts~\cite{PhysRevLett.120.140404}. Consequently, the structure of the incoherent power spectra finds its explanation in the roots of the denominators in $A_1^c(E/2-\omega)$ and $A_2^c(E/2-\omega)$, which are given by
\begin{align}
\omega_1^c=&E/2-\eta_c/2-i\tilde{\Gamma}_c/2
\end{align}
for $A_1^c(\omega)$, and 
\begin{align}
\omega_2^c=&E/2-\eta_c/2+i\tilde{\Gamma}_c/2
\end{align}
for $A_2^c(\omega)$. The respective values corresponding to the given parameters in Fig.~\ref{fig2}  are listed in the Table~\ref{tab:table1}.  The incoherent power spectra differ in the transmission and reflection for the separate and braided configuration. This results from the exchange of atomic excitations in the atoms for incident photons moving in opposite directions, owing to the parity symmetry $\hat{P}\hat{\sigma}_j\hat{P}^\dagger=\hat{\sigma}_{3-j}$. This behavior is also consistent with the small atoms system~\cite{PhysRevA.91.053845,PhysRevA.88.043806,science.1244324}. Conversely, in the nested configuration, their incoherent power spectra remain the same because of the unchanged atomic excitations for incident photons moving in both directions, where the parity symmetry is represented as $\hat{P}\hat{\sigma}_j\hat{P}^\dagger=\hat{\sigma}_{j}$.
\begin{table}[hbt]
\caption{\label{tab:table1}
The numerical values of $\omega_1^c(\omega)$ and $\omega_2^c(\omega)$ (in units of $\Gamma$) for the three configurations.}
\begin{ruledtabular}
\begin{tabular}{ccc}
 & $\phi_1=0.5\pi$, $\phi_2=0.25\pi$& $\phi_1=0.25\pi$, $\phi_2=0.85\pi$ \\
\hline
Separate&$\omega_1^s=101.7-0.3i$ & $\omega_1^s=100.2-0.08i$  \\
&$\omega_2^s=100.3-1.7i$ & $\omega_2^s=101.2-3.3i$  \\ \hline
Braided& $\omega_1^b=101.7-0.3i$ & $\omega_1^b=100.2-0.08i$  \\
&$\omega_2^b=99.7-0.3i$ & $\omega_2^b=99.2-0.014i$  \\ \hline
Nested &$\omega_1^n=98.4-1.4i$ & $\omega_1^n=99-0.64i$  \\
&$\omega_2^n=101.6-0.56i$ & $\omega_2^n=100.6-0.013i$  \\            
\end{tabular}
\end{ruledtabular}
\label{table1}
\end{table}

Furthermore, the total inelastic flux is defined as
\begin{align}
F^c(k)=&\int d\omega S^{c}_\text{incoh}(\omega)\nonumber\\
=&\frac{2}{\pi}\int dx\left[\abs{B_{k_1k_2}^{RR,c}(x)}^2+\abs{B_{k_1k_2}^{LL,c}(x)}^2\right].
\end{align}
This provides a measure of the overall strength of correlations and a direct measurement of the bound-state term. When $k_1=k_2=k$, after integration, the expression for the total inelastic flux can be obtained as
\begin{align}
F^{c}(k)&=\frac{8}{\pi}\Bigg[\frac{\abs{Z_1^c}^2+\abs{Z_3^c}^2}{i\eta_c^*-i\eta_c+\tilde{\Gamma}_c^*+\tilde{\Gamma}_c}+\frac{\abs{Z_2^c}^2+\abs{Z_4^c}^2}{i\eta_c^*-i\eta_c-\tilde{\Gamma}_c^*-\tilde{\Gamma}_c}\nonumber\\&+\frac{Z_1^{c*}Z_2^c+Z_3^{c*}Z_4^c}{i\eta_c^{*}-i\eta_c+\tilde{\Gamma}_c^{c*}-\tilde{\Gamma}_c}
+\frac{Z_1^cZ_2^{c*}+Z_3^cZ_4^{c*}}{i\eta_c^*-i\eta_c-\tilde{\Gamma}_c^*+\tilde{\Gamma}_c}\Bigg].
\end{align}

\begin{figure}[hbt]
\centering\includegraphics[width=8cm,keepaspectratio,clip]{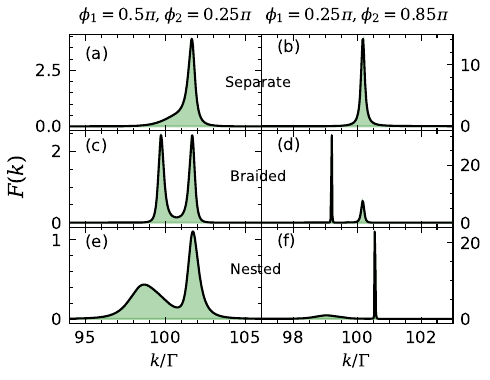}
\caption{The total inelastic flux in the three different configurations as a function of $k$ with different values of $\phi_1$ and $\phi_2$. The first row corresponds to the separate case, the second row corresponds to the braided case, and the third row corresponds to the nested case. The other parameter is $\omega_0=100\Gamma$.}
\label{fig3}
\end{figure}

The total inelastic flux for the three configurations $F^{c}(k)$ as a function of the incident frequency $k$ is shown in Fig.~\ref{fig3}. A large value of $F^c(k)$ indicates strong correlation effects, since the incoherent scattering arises from the correlation of the bound state. Therefore, the peak value indicates the strongest correlation, and the corresponding $k_\text{peak}$ represents the optimal incident frequency to obtain photon-photon correlation. The shape of $F^c(k)$ varies with the accumulated phase shifts $\phi_1$ and $\phi_2$. The position and width of the peaks can be explained by the poles of the system, which correspond to the roots of $D^c$. Denoting the poles as $z=\tilde{\omega}-i\tilde{\Gamma}$, $\tilde{\omega}$ represents the eigenfrequency, and $\tilde{\Gamma}$ denotes the collective decay rate~\cite{Dinc2019exactmarkoviannon,Qiu2023}. The position of the peak aligns with the eigenfrequency $\tilde{\omega}$, while its width is determined by $\tilde{\Gamma}$.

To validate our analytical results for the incoherent power spectra and total inelastic flux, we employ the master equation approach, which involves tracing out the 1D bosonic modes in the waveguide~\cite{PhysRevLett.120.140404,PhysRevA.88.043806}. Moreover, we consider a weak probe field that contains multi-photon components, extending beyond the single-photon limit~\cite{Jia2022}. However, it is important to note that the dominant processes primarily involve two photons. A detailed derivation of these calculations is presented in Appendix~\ref{appendix:B}. The presence of waveguide modes induces various effects on the giant atoms, including frequency shifts, exchange interactions, individual decay, and collective decay. All these parameters depend on the accumulated phase shift. The output field within the waveguide consists of both a coherent term and an incoherent term, similar to the decomposition observed in two-photon wavefunction described by Eq.~\eqref{eqn:two-photon_amp}. Consequently, the incoherent power spectra correspond to the collective resonance fluorescence emitted from the giant atoms. In their eigenstate representation, the spectrum manifests as the sum of the resonance fluorescence from the eigenstates. Thus, the positions and widths of the peaks listed in Table~\ref{tab:table1} align with the eigenfrequency and its effective dissipation rate, as shown in Fig.~\ref{fig6}. Furthermore, the plots depicting the incoherent power spectra and total inelastic flux in Fig.~\ref{fig7}, obtained through numerical simulation of the master equation, are consistent with the results derived from the analytical expression of the two-photon wave function.

\section{Second-order correlation function}
\label{sec6}
\begin{figure}[hbt]
\centering\includegraphics[width=8cm,keepaspectratio,clip]{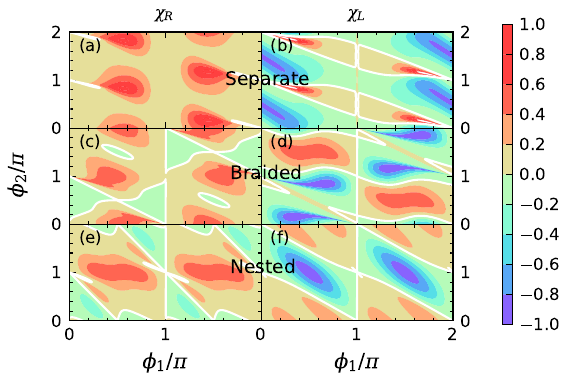}
\caption{Second-order differential correlation functions in the three different configurations as functions of $\phi_1$ and $\phi_2$. The first row corresponds to the separate configuration, the second row corresponds to the braided configuration, and the third row corresponds to the nested configuration. Moreover, the first column represents the transmission, and the second column represents the reflection. The white lines indicate that the differential correlation functions equal to zero. The other parameters are $k=\omega_0=100\Gamma$.}
\label{fig4}
\end{figure}

Next, we utilize the second-order correlation function to demonstrate the spatial interaction between photons~\cite{loudon2000quantum}. The second-order correlation functions of the transmitted and reflected fields ($x_1>d/2$, $x_2>d/2$ and $x=x_2-x_1$) are defined as follows:
\begin{align}
{G_\alpha^c}^{(2)}(x)=&\bra{\Psi_2^c}\hat{a}^\dagger_\alpha(x_1)\hat{a}^\dagger_\alpha(x_2)\hat{a}_\alpha(x_2)\hat{a}_\alpha(x_1)\ket{\Psi_2^c}\nonumber\\
&=2\abs{f_{\alpha\alpha}^c(x_1,x_2)}^2.
\end{align}
This correlation function represents the probability of detecting a photon at $x_2$ after detecting the first one at $x_1$. The expression is directly proportional to the rate at which two photons are transmitted or reflected, and is determined by the interference between the plane-wave term and the bound-state term. In order to briefly illustrate the effect of the bound state, we examine the second-order differential correlation function~\cite{PhysRevLett.101.203602,PhysRevA.85.041801,PhysRevA.87.025803,PhysRevA.100.053857}, which is the difference between the probability of two-photon detection and the independent single-photon detection when $x=0$. Concretely, under the condition that $k_1=k_2=k$  the differential correlation function are
\begin{align}
\chi_R^c=2\pi^2\abs{f_{RR}^c(0)}^2-\abs{t_4^c(k)}^4
\end{align}
for the transmitted field and
\begin{align}
\chi_L^c=2\pi^2\abs{f_{LL}^c(0)}^2-\abs{r_1^c(k)}^4
\label{eqn:chi_L}
\end{align}
for the reflected field. If $\chi_R>0$, it indicates that the bound state enhances the transmission of two photons, resulting in a phenomenon known as photon-induced tunneling, which serves as a signature of photon bunching. Conversely, if $\chi_R<0$, it implies that the bound state can suppress the transmission of two photons, leading to photon blockade~\cite{PhysRevLett.107.223601,PhysRevA.85.043832}. The second-order differential correlation functions for the three configurations are numerically plotted in Fig.~\ref{fig4}. 

In the single giant atom, the photon correlation can be enhanced by adjusting the phase shift, but it is unable to switch between bunching and antibunching~\cite{PhysRevA.108.053718}. This limitation arises because a single two-level atom can absorb only one photon at a time and cannot emit two photons simultaneously. Hence, in the case of reflection from a single giant atom, the second-order correlation function $g^{(2)}(0)=0$. However, this constraint can be overcome by incorporating additional two-level atoms. The presence of multiple two-level atoms enables the possibility of one photon being absorbed by the first atom while the other photon propagates to the subsequent atom and gets reflected, thereby triggering the stimulated emission of the first photon. As a result, the probability of two photons being emitted together is not completely suppressed. This phenomenon becomes even more pronounced when two giant atoms are present. In the case of transmission for the two giant atoms, we find that the transmitted photons exhibit bunching behavior when the atoms are in the separate configuration, similar to small atoms. However, the transmitted photons can display either bunching or antibunching behavior by adjusting the phase shifts $\phi_1$ and $\phi_2$ in the braided and nested cases. As for the reflection, the reflected photons can exhibit either bunching or antibunching behavior by adjusting the phase shifts $\phi_1$ and $\phi_2$ in all three cases.

Alternatively, the statistics of two photons can be interpreted by considering the interplay between a coherent state and a squeezed state~\cite{https://doi.org/10.1002/lpor.201900279,LópezCarreño2018,PhysRevLett.108.093602}. The wave function presented in Eq.~\eqref{eqn:two-photon_amp}, comprises a coherent state arising from the coherent scattering and a squeezed-like state originated from the bound state (under the condition of $k_1=k_2=k$). The combination of these states gives rise to bunching and antibunching phenomena, which are closely related to the phases associated with the coherent and squeezing components. Remarkably, the relative phase can be adjusted through the accumulated phase shifts between the giant atoms. This perspective can also be numerically validated using the master equation approach in Appendix~\ref{appendix:B}. In the eigenstate representation shown by Fig.~\ref{fig6}, the system exhibits behavior reminiscent of two independent coherently driven two-level atoms in the Heitler regime. While each effective two-level atom provides antibunched resonance fluorescence, the total output field, expressed as a phase-dependent sum of output field from each effective two-level atom, and the phase itself plays a key role in tuning the photon correlations. These characteristics can be referred to as unconventional statistic features~\cite{https://doi.org/10.1002/lpor.201900279}. The numerical simulations of the two-photon differential correlation function in Fig.~\ref{fig8} exhibit a strong resemblance to those derived from our analytical results. This validation provides further support for our analytical results and reinforces the underlying mechanism responsible for the photon statistics.

\begin{figure}[hbt]
\centering\includegraphics[width=8.5cm,keepaspectratio,clip]{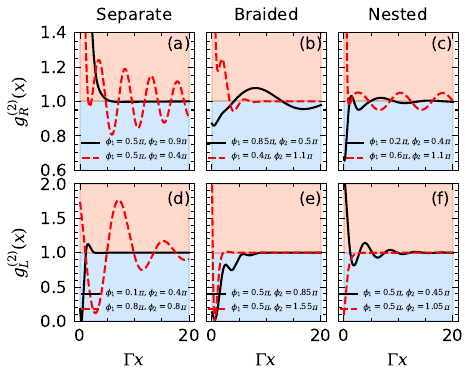}
\caption{Normalized second-order correlation functions in the three different configurations as a function of $x$ with different values of $\phi_1$ and $\phi_2$. The first row corresponds to the transmission, and the second row corresponds to the reflection. Additionally, the first column represents the separate case, the second column represents the braided case, and the third column represents the nested case. The other parameters are $k=\omega_0=100\Gamma$.}
\label{fig5}
\end{figure}

The differential correlation function provides a clear measure of the probability of generating two photons simultaneously. However, it is less sensitive to single-photon transmission and reflection. To address this, it is helpful to introduce the normalized second-order correlation function~\cite{Y.Fang@EPJ2014,PhysRevA.107.023704,PhysRevA.101.053812,science.aao7293}
\begin{align}
{g_\alpha^c}^{(2)}(x)=\frac{{G^c_\alpha}^{(2)}(x)}{\abs{_\alpha\langle x_1\ket{\Phi_1^c(k_1)}_R}^2\abs{_\alpha\langle x_2\ket{\Phi_1^c(k_2)}_R}^2}.
\end{align}
This function is normalized by the single-photon transmission and reflection probabilities. After performing calculations, the normalized second-order correlation functions in transmission and reflection can be expressed in the form
\begin{align}
{g_R^c}^{(2)}(x)=&\bigg\vert1+\frac{Z_1^c}{t_4^c(k_1)t_4^c(k_2)}e^{(i\eta_c-\tilde{\Gamma}_c)\abs{x}/2}\nonumber\\&+\frac{Z_2^c}{t_4^c(k_1)t_4^c(k_2)}e^{(i\eta_c+\tilde{\Gamma}_c)\abs{x}/2}\bigg\vert^2,\nonumber\\
{g_L^c}^{(2)}(x)=&\bigg\vert1+\frac{Z_3^c}{r_1^c(k_1)r_1^c(k_2)}e^{(i\eta_c-\tilde{\Gamma}_c)\abs{x}/2}\nonumber\\&+\frac{Z_4^c}{r_1^c(k_1)r_1^c(k_2)}e^{(i\eta_c+\tilde{\Gamma}_c)\abs{x}/2}\bigg\vert^2.
\end{align}
These correlation functions are shown in Fig.~\ref{fig5}. Here we choose the frequency of the input field to be resonant with the atomic transition frequency, i.e., $k_1=k_2=k=\omega_0$. It should be noted that, consistent with Fig.~\ref{fig4}, in Fig.~\ref{fig5}(a), the transmission correlations both exhibit bunching behavior [${g_R^s}^{(2)}(0)>0$] in the separate configuration. However, in the braided and nested configurations shown in Figs.~\ref{fig5}(b) and \ref{fig5}(c), the initial bunching ${g_R^{b/n}}^{(2)}(0)>0$ and antibunching ${g_R^{b/n}}^{(2)}(0)<0$ can be manipulated by adjusting the phase shifts $\phi_1$ and $\phi_2$. As for reflection, the correlation can either display bunching or antibunching behavior in all three configurations by adjusting the phase shifts. 

Besides, it is important to note that the initial value ${g_\alpha^c}^{(2)}(x)$ cannot predict the overall photon-photon correlation due to the complex nature of the function. This complexity arises from the beating between the incident frequency and the two eigenfrequencies $(\eta_c\pm i\tilde{\Gamma}_c)/2$, which correspond to the complete set of collective decay rates~\cite{Dinc2019exactmarkoviannon, Qiu2023}. The long-distance behavior is determined by the most sub-radiant pole~\cite{PhysRevA.91.053845}. For example, in the separate configuration shown in Fig.~\ref{fig5}(a), the imaginary of the most sub-radiant pole is $0.7$ for $\phi_1=0.5\pi$ and $\phi_2=0.9\pi$ (black solid line), while it is $0.05$ for $\phi_1=0.5\pi$ and $\phi_2=0.4\pi$ (red-dashed line), which exhibits the long-distance oscillation. A similar analysis can be applied to the other sub-figures as well. In the limit of large $x$, the contribution of the bound state becomes negligible, and the second-order correlation function approaches to $1$.

\section{Conclusions and Discussions}
\label{conclusion}
In conclusion, we have instigated the two-photon scattering processes involving two giant atoms coupled to a 1D waveguide. We study three different configurations: separate, braided, and nested, using the LS formalism. The approach enables us to obtain analytical expressions for the two-photon interacting scattering wave functions under the Markovian approximation. Based on our analytical results, we derived the incoherent power spectrum, which arises from the correlation of the bound state and characterizes the generation of correlated photon pairs. Importantly, we demonstrated that the incoherent power spectrum can be effectively tuned by adjusting the accumulated phase shifts. Furthermore, we analyzed the total flux as a measure of photon-photon correlation, showing that it can be modified by the phase shifts and explained by the poles of the system. The second-order correlation function provides a direct measure of photon-photon correlation, and our analysis revealed that the accumulated phase shifts can be effectively utilized to qualitatively tune the photon-photon correlation. This includes manipulating either initial bunching or initial antibunching behavior in the transmission and reflection. Lastly, we found that the long-distance evolution of the second-order correlation is possible from the most sub-radiant poles. This work offers possibilities for generating tunable nonclassical photon source, which may have potential applications in the construction of quantum networks based on the giant-atom waveguide-QED systems.

The effect of waveguide loss has been completely neglected in the present study. Let us now consider the potential impact of such losses. In waveguide systems, the losses associated with light propagation are typically very low, especially when considering high-quality materials and careful design techniques~\cite{https://doi.org/10.1002/lpor.201300183,Horikawa2016}. The primary source of losses usually comes from external components, such as circulators used in the unidirectional waveguides~\cite{PhysRevX.7.041043}. In our case, since we consider a bidirectional waveguide, there is no need for lossy circulators. To account for waveguide loss theoretically, they can be modelled by introducing a beam-splitter mixing term, where $\hat{B}_\text{out}^{(m)}=\sqrt{\eta}\hat{b}_\text{out}^{(m)}+\sqrt{1-\eta}\hat{b}_v$. Here, $\eta$ represents the transmission efficiency, denoting the fraction of the input light transmitted through the waveguide, and $\hat{b}_v$ is the uncorrelated noise. If the noise is assumed to be the vacuum noise, then the incoherent power spectrum would become $\eta S_\text{incoh}(\omega)$, the differential correlation function would become $\eta^2\chi^{(m)}$, and the normalized second-order correlation would remain the same.

\begin{acknowledgments}
This work was supported by the National Natural Science Foundation of China (under Grants No. 1150403, No. 61505014, and No. 12174139.)
\end{acknowledgments}

\appendix
\section{Derivation of bound-state terms in three configurations}
\label{appendix}
In this Appendix, we present the analytical results of the bound-state terms in the separate, braided, and nested configurations.
\subsection{Separate-coupling case}
The elements of Green's functions can be obtained by performing a double integral using standard contour integral techniques. In the separate configuration, we have explicitly derived the elements as follows:
\begin{align}
&G_{11}^s=\frac{2\eta_s^2+\tilde{\Gamma}_s^2}{2\eta_s\left(\eta_s^2+\tilde{\Gamma}_s^2\right)},\hspace{5pt}
G_{12}^s=\frac{-\tilde{\Gamma}_s^2}{2\eta_s\left(\eta_s^2+\tilde{\Gamma}_s^2\right)},\nonumber\\
&G_{1,s}^{RR}(x_1,x_2)=-C_s\left(\Upsilon_1^se^{-\tilde{\Gamma}x/2}+\Upsilon_2^se^{\tilde{\Gamma}x/2}\right),\nonumber\\
&G_{2,s}^{RR}(x_1,x_2)=-C_s\left(\Upsilon_3^se^{-\tilde{\Gamma}x/2}+\Upsilon_4^se^{\tilde{\Gamma}x/2}\right),\nonumber\\
&G_{1,s}^{RL}(x_1,-x_2)=-C_s\left(\Upsilon_3^se^{-\tilde{\Gamma}x/2}-\Upsilon_4^se^{\tilde{\Gamma}x/2}\right),\nonumber\\
&G_{2,s}^{RL}(x_1,-x_2)=-C_s\left(\Upsilon_1^se^{-\tilde{\Gamma}x/2}-\Upsilon_2^se^{\tilde{\Gamma}x/2}\right),
\end{align}
where
\begin{align}
\eta_s=&E-2\omega_0+2i\Gamma(1+e^{i\phi_1}),\nonumber\\
C_s=&\Gamma\cos^2\frac{\phi_1}{2}e^{i(\eta_s x/2+Ex_c)},\nonumber\\
\Upsilon_1^s=&\frac{2\eta_s+i\tilde{\Gamma}_s-2i\Gamma(1+\cos\phi_1)}{2\eta_s(\eta_s+i\tilde{\Gamma}_s)}\left[e^{i(\phi_1+\phi_2)}+1\right],\nonumber\\
\Upsilon_2^s=&\frac{2\eta_s-i\tilde{\Gamma}_s-2i\Gamma(1+\cos\phi_1)}{2\eta_s(\eta_s-i\tilde{\Gamma}_s)}\left[e^{i(\phi_1+\phi_2)}-1\right],\nonumber\\
\Upsilon_3^s=&\frac{2\eta_s+i\tilde{\Gamma}_s\left[1-e^{i(\phi_1+\phi_2)}\right]}{2\eta_s(\eta_s+i\tilde{\Gamma}_s)}\left[e^{-i(\phi_1+\phi_2)}+1\right],\nonumber\\
\Upsilon_4^s=&\frac{2\eta_s-i\tilde{\Gamma}_s(1+e^{i(\phi_1+\phi_2)})}{2\eta_s(\eta_s-i\tilde{\Gamma}_s)}\left[e^{-i(\phi_1+\phi_2)}-1\right].
\end{align}

According to the parity symmetry $\hat{P}\hat{\sigma}_j\hat{P}^\dagger=\hat{\sigma}_{3-j}$, the following equalities hold: $G_{21}^s=G_{12}^s$, $G_{22}^s=G_{11}^s$, $G_{1,s}^{LL}(-x_1,-x_2)=G_{2,s}^{RR}(x_1,x_2)$, and $G_{2,s}^{LL}(-x_1,-x_2)=G_{1,s}^{RR}(x_1,x_2)$. Additionally, it can be proven that $G_{2,s}^{RL}(x_1,-x_2)=G_{1,s}^{LR}(-x_1,x_2)$, and $G_{2,s}^{LR}(-x_1,x_2)=G_{1,s}^{RL}(x_1,-x_2)$. Then, following the Eq.~\eqref{eqn:LS}, the bound-state terms in transmission and reflection amplitudes can be expressed in the form
\begin{align}
B_{k_1k_2}^{RR,s}(x)=&Z_1^se^{(i\eta_s-\tilde{\Gamma}_s)\abs{x}/2}+Z_2^se^{(i\eta_s+\tilde{\Gamma}_s)\abs{x}/2},\nonumber\\
B_{k_1k_2}^{LL,s}(x)=&Z_3^se^{(i\eta_s-\tilde{\Gamma}_s)\abs{x}/2}+Z_4^se^{(i\eta_s+\tilde{\Gamma}_s)\abs{x}/2},
\end{align}
where the coefficients are
\begin{align}
Z_1^s=&\frac{\sqrt{2}\pi\Gamma\cos^2\frac{\phi_1}{2}}{{G_{11}^s}^2-{G_{12}^s}^2}\Big[e_{1R}^s(k_1)e_{1R}^s(k_2)(\Upsilon_1^sG_{11}^s-\Upsilon_3^sG_{12}^s)\nonumber\\&+e_{2R}^s(k_1)e_{2R}^s(k_2)(\Upsilon_3^sG_{11}^s-\Upsilon^s_1G_{12}^s)\Big],\nonumber\\
Z_2^s=&\frac{\sqrt{2}\pi\Gamma\cos^2\frac{\phi_1}{2}}{{G_{11}^s}^2-{G_{12}^s}^2}\Big[e_{1R}^s(k_1)e_{1R}^s(k_2)(\Upsilon_2^sG_{11}^s-\Upsilon_4^sG_{12}^s)\nonumber\\&+e_{2R}^s(k_1)e_{2R}^s(k_2)(\Upsilon_4^sG_{11}^s-\Upsilon_2^sG_{12}^s)\Big],\nonumber\\
Z_3^s=&\frac{\sqrt{2}\pi\Gamma\cos^2\frac{\phi_1}{2}}{{G_{11}^s}^2-{G_{12}^s}^2}\Big[e_{1R}^s(k_1)e_{1R}^s(k_2)(\Upsilon_3^sG_{11}^s-\Upsilon_1^sG_{12}^s)\nonumber\\&+e_{2R}^s(k_1)e_{2R}^s(k_2)(\Upsilon_1^sG_{11}^s-\Upsilon_3^sG_{12}^s)\Big],\nonumber\\
Z_4^s=&\frac{\sqrt{2}\pi\Gamma\cos^2\frac{\phi_1}{2}}{{G_{11}^s}^2-{G_{12}^s}^2}\Big[e_{1R}^s(k_1)e_{1R}^s(k_2)(\Upsilon_4^sG_{11}^s-\Upsilon_2^sG_{12}^s)\nonumber\\&+e_{2R}^s(k_1)e_{2R}^s(k_2)(\Upsilon_2^sG_{11}^s-\Upsilon_4^sG_{12}^s)\Big].
\end{align}

\subsection{Braided-coupling case}
Following the same procedure, in the braided-coupling configuration, we can also express the bound-state terms as
\begin{align}
B_{k_1k_2}^{RR,b}(x)=&Z_1^be^{(i\eta_b-\tilde{\Gamma}_b)\abs{x}/2}+Z_2^be^{(i\eta_b+\tilde{\Gamma}_b)\abs{x}/2},\nonumber\\
B_{k_1k_2}^{LL,b}(x)=&Z_3^be^{(i\eta_b-\tilde{\Gamma}_b)\abs{x}/2}+Z_4^be^{(i\eta_b+\tilde{\Gamma}_b)\abs{x}/2}.
\end{align}
Here, the coefficients are
\begin{align}
Z_1^b=&\frac{\sqrt{2}\pi\Gamma\cos^2\left(\frac{\phi_1+\phi_2}{2}\right)}{{G_{11}^b}^2-{G_{12}^b}^2}\Big[e_{1R}^b(k_1)e_{1R}^b(k_2)(\Upsilon_1^bG_{11}^b\nonumber\\&-\Upsilon_3^bG_{12}^b)+e_{2R}^b(k_1)e_{2R}^b(k_2)(\Upsilon_3^bG_{11}^b-\Upsilon_1^bG_{12}^b)\Big],\nonumber\\
Z_2^b=&\frac{\sqrt{2}\pi\Gamma\cos^2\left(\frac{\phi_1+\phi_2}{2}\right)}{{G_{11}^b}^2-{G_{12}^b}^2}\Big[e_{1R}^b(k_1)e_{1R}^b(k_2)(\Upsilon_2^bG_{11}^b\nonumber\\&-\Upsilon_4^bG_{12}^b)+e_{2R}^b(k_1)e_{2R}^b(k_2)(\Upsilon_4^bG_{11}^b-\Upsilon_2^bG_{12}^b)\Big],\nonumber\\
Z_3^b=&\frac{\sqrt{2}\pi\Gamma\cos^2\left(\frac{\phi_1+\phi_2}{2}\right)}{{G_{11}^b}^2-{G_{12}^b}^2}\Big[e_{1R}^b(k_1)e_{1R}^b(k_2)(\Upsilon_3^bG_{11}^b\nonumber\\&-\Upsilon_1^bG_{12}^b)+e_{2R}^b(k_1)e_{2R}^b(k_2)(\Upsilon_1^bG_{11}^b-\Upsilon_3^bG_{12}^b)\Big],\nonumber\\
Z_4^b=&\frac{\sqrt{2}\pi\Gamma\cos^2\left(\frac{\phi_1+\phi_2}{2}\right)}{{G_{11}^b}^2-{G_{12}^b}^2}\Big[e_{1R}^b(k_1)e_{1R}^b(k_2)(\Upsilon_4^bG_{11}^b\nonumber\\&-\Upsilon_2^bG_{12}^b)+e_{2R}^b(k_1)e_{2R}^b(k_2)(\Upsilon_2^bG_{11}^b-\Upsilon_4^bG_{12}^b)\Big],\nonumber\\
\end{align}
where the parameters involved in these expressions are
\begin{align}
\eta_b=&E-2\omega_0+2i\Gamma\left[1+e^{i(\phi_1+\phi_2)}\right],\nonumber\\
G_{11}^b=&\frac{2\eta_b^2+\tilde{\Gamma}_b^2}{2\eta_b\left(\eta_b^2+\tilde{\Gamma}_b^2\right)},\hspace{5pt}
G_{12}^b=\frac{-\tilde{\Gamma}_b^2}{2\eta_b\left(\eta_b^2+\tilde{\Gamma}_b^2\right)},\nonumber\\
\Upsilon_1^b=&\frac{2\eta_b+i\tilde{\Gamma}_b-2i\Gamma(1+e^{i\phi_2}\cos\phi_1)}{2\eta_b(\eta_b+i\tilde{\Gamma}_b)}\left(e^{i\phi_1}+1\right),\nonumber\\
\Upsilon_2^b=&\frac{2\eta_b-i\tilde{\Gamma}_b-2i\Gamma(1+e^{i\phi_2}\cos\phi_1)}{2\eta_b(\eta_b-i\tilde{\Gamma}_b)}\left(e^{i\phi_1}-1\right),\nonumber\\
\Upsilon_3^b=&\frac{2\eta_b+i\tilde{\Gamma}_b(1-e^{i\phi_1})}{2\eta_b(\eta_b+i\tilde{\Gamma}_b)}\left(e^{-i\phi_1}+1\right),\nonumber\\
\Upsilon_4^b=&\frac{2\eta_b-i\tilde{\Gamma}_b(1+e^{i\phi_1})}{2\eta_b(\eta_b-i\tilde{\Gamma}_b)}\left(e^{-i\phi_1}-1\right).
\end{align}

\subsection{Nested-coupling case}
Following the same procedure, the bound-state terms in the nested-coupling configuration can be expressed as
\begin{align}
B_{k_1k_2}^{RR,n}(x)=&Z_1^ne^{(i\eta_n-\tilde{\Gamma}_n)\abs{x}/2}+Z_2^ne^{(i\eta_n+\tilde{\Gamma}_n)\abs{x}/2},\nonumber\\
B_{k_1k_2}^{LL,n}(x)=&Z_3^ne^{(i\eta_n-\tilde{\Gamma}_n)\abs{x}/2}+Z_4^ne^{(i\eta_n+\tilde{\Gamma}_n)\abs{x}/2},
\end{align}
where the coefficients are
\begin{align}
Z_1^n=&\frac{-i2\sqrt{2}\pi\Gamma}{G_{11}^nG_{22}^n-{G_{12}^n}^2}\Big[e_{1R}^n(k_1)e_{1R}^n(k_2)(\Upsilon_1^nG_{22}^n-\Upsilon_3^nG_{12}^n)\nonumber\\&+e_{2R}^n(k_1)e_{2R}^n(k_2)(\Upsilon_3^nG_{11}^n-\Upsilon_1^nG_{12}^n)\Big],\nonumber\\
Z_2^n=&\frac{i2\sqrt{2}\pi\Gamma}{G_{11}^nG_{22}^n-{G_{12}^n}^2}\Big[e_{1R}^n(k_1)e_{1R}^n(k_2)(\Upsilon_2^nG_{22}^n-\Upsilon_4^nG_{12}^n)\nonumber\\&+e_{2R}^n(k_1)e_{2R}^n(k_2)(\Upsilon_4^nG_{11}^n-\Upsilon_2^nG_{12}^n)\Big].
\end{align}
Here, $Z_3^n=Z_1^n$ and $Z_4^n=Z_2^n$ due to the fact that the atoms remain unchanged for the right-moving and left-moving incident photons, i.e., $\hat{P}\hat{\sigma}_j\hat{P}^\dagger=\hat{\sigma}_j$. The parameters involved in these expressions are given as
\begin{align}
\eta_n=&E-2\omega_0+2i\Gamma\left[1+\cos\phi_1e^{i(\phi_1+\phi_2)}\right],\nonumber\\
G_{11}^n=&\frac{\eta_n^2+i\Gamma\eta_n e^{i\phi_2}\left(1-e^{i2\phi_1}\right)+2\Gamma^2e^{i2\phi_1}\left(1+e^{i\phi_2}\right)^2}{\eta_n(\eta_n^2+\tilde{\Gamma}_n^2)},\nonumber\\
G_{22}^n=&\frac{\eta_n^2-i\Gamma\eta_n e^{i\phi_2}\left(1-e^{i2\phi_1}\right)+2\Gamma^2e^{i2\phi_1}\left(1+e^{i\phi_2}\right)^2}{\eta_n(\eta_n^2+\tilde{\Gamma}_n^2)},\nonumber\\
G_{12}^n=&\frac{-2\Gamma^2e^{i2\phi_1}\left(1+e^{i\phi_2}\right)^2}{\eta_n(\eta_n^2+\tilde{\Gamma}_n^2)},
\end{align}
and
\begin{align}
\Upsilon_1^n=&\Big\{\Gamma^2\cos^2\left(\frac{\phi_2}{2}\right)(e^{i2\phi_1}-1)
(e^{i\phi_2}+1)\nonumber\\&-\eta_n\Gamma\sin\phi_1[\cos\phi_1+\cos(\phi_1+\phi_2)]\nonumber\\&+\eta_n(\omega_0-\lambda_2^n)\cos^2\left(\phi_1+\frac{\phi_2}{2}\right)\Big\}\nonumber\\&/[\eta_n\tilde{\Gamma}_n(\eta_n+i\tilde{\Gamma}_n)],\nonumber\\
\Upsilon_2^n=&\Big\{\Gamma^2\cos^2\left(\frac{\phi_2}{2}\right)(e^{i2\phi_1}-1)(e^{i\phi_2}+1)\nonumber\\&-\eta_n\Gamma\sin\phi_1[\cos\phi_1+\cos(\phi_1+\phi_2)]\nonumber\\&+\eta_n(\omega_0-\lambda_1^n)\cos^2\left(\phi_1+\frac{\phi_2}{2}\right)\Big\}\nonumber\\&/[\eta_n\tilde{\Gamma}_n(\eta_n-i\tilde{\Gamma}_n)],\nonumber\\
\Upsilon_3^n=&\cos^2\left(\frac{\phi_2}{2}\right)\frac{\eta_n(\omega_0-\lambda_2^n)-\Gamma^2(e^{i2\phi_1}-1)
(e^{i\phi_2}+1)}{\eta_n\tilde{\Gamma}_n(\eta_n+i\tilde{\Gamma}_n)},\nonumber\\
\Upsilon_4^n=&\cos^2\left(\frac{\phi_2}{2}\right)\frac{\eta_n(\omega_0-\lambda_1^n)-\Gamma^2(e^{i2\phi_1}-1)
(e^{i\phi_2}+1)}{\eta_n\tilde{\Gamma}_n(\eta_n-i\tilde{\Gamma}_n)}.
\end{align}
Here, $\lambda_{1,2}^n$ are the roots of $D^n$ and given by
\begin{align}
\lambda_{1,2}^n=&\omega_0-i\frac{\Gamma}{2}\left[2+e^{i\phi_2}+e^{i(2\phi_1+\phi_2)}\right]\pm i\frac{\tilde{\Gamma}_n}{2}.
\end{align}

\section{Analysis of incoherent power spectrum and two-photon differential correlation functions using the master equation approach}
\label{appendix:B}
In this Appendix, we validate the photon scattering processes discussed in Sec.~\ref{sec5} using the master equation approach. Our treatment considers the weak coherent field as a probe field, which contains multi-photon components. This allows for scattering beyond the single-photon limit, with two-photon processes being predominant. 

To derive the master equation for the density operator $\hat{\rho}$ of the double giant atoms, we trace out the continuum of bosonic modes in the waveguide and work in a frame rotating with the driving frequency. The resulting master equation is given by~\cite{PhysRevA.104.063712,PhysRevLett.120.140404}
\begin{align}
\frac{d}{dt}\hat{\rho}=&-i[\hat{H}_\text{dr},\hat{\rho}]+\sum_j\Gamma_j\mathcal{D}[\hat{\sigma}_j^-]\hat{\rho}\nonumber\\
&+\Gamma_{12}\sum_{j\neq j^\prime}\left(\hat{\sigma}_j^-\hat{\rho}\hat{\sigma}_{j^\prime}^+-\frac{1}{2}\{\hat{\sigma}_j^+\hat{\sigma}_{j^\prime}^-,\hat{\rho}\}\right),
\end{align}
where
\begin{align}
\hat{H}_\text{dr}=&\sum_{j}\Delta_{L_j}\hat{\sigma}_j^+\hat{\sigma}_j^-+g_{12}\left(\hat{\sigma}_1^+\hat{\sigma}_2^-+\hat{\sigma}_2^+\hat{\sigma}_1^-\right)\nonumber\\&-\frac{i}{2}\sum_j\left(\Omega_j\hat{\sigma}_j^+-\text{H.c.}\right).
\end{align}
Here, $\mathcal{D}[\hat{O}]\hat{\rho}=\hat{O}\hat{\rho}\hat{O}^\dagger-\{\hat{O}^\dagger\hat{O},\hat{\rho}\}/2$ is the Lindblad operator, and $\Omega_j$ is the Rabi frequency of the $j$th atom. For the separate giant atoms, we have
\begin{align}
\Delta_{L_1}=&\Delta_{L_2}=\Gamma\sin\phi_1,\hspace{5pt} \Gamma_1=\Gamma_2=2\Gamma\left(1+\cos\phi_1\right),\nonumber\\
g_{12}=&\frac{\Gamma}{2}\left[\sin\phi_2+2\sin\left(\phi_1+\phi_2\right)+\sin\left(2\phi_1+\phi_2\right)\right],\nonumber\\
\Gamma_{12}=&\Gamma\left[\cos\phi_2+2\cos\left(\phi_1+\phi_2\right)+\cos\left(2\phi_1+\phi_2\right)\right],\nonumber\\
\Omega_1=&\sqrt{2\Gamma}\alpha\left(1+e^{i\phi_1}\right),\nonumber\\
\Omega_2=&\sqrt{2\Gamma}\alpha\left[e^{i(\phi_1+\phi_2)}+e^{i(2\phi_1+\phi_2)}\right],
\end{align}
where $\alpha$ is the strength of the weak coherent drive. Using the input-output relation, the reflection and transmission fields are defined as
\begin{align}
\hat{b}_\text{out}^{(r)}=&\sqrt{\frac{\Gamma}{2}}(1+e^{i\phi_1})\left[\hat{\sigma}_1^-+e^{i(\phi_1+\phi_2)}\hat{\sigma}_2^-\right],\nonumber\\
\hat{b}_\text{out}^{(t)}=&\alpha e^{i(2\phi_1+\phi_2)}+\sqrt{\frac{\Gamma}{2}}(1+e^{i\phi_1})\left[e^{i(\phi_1+\phi_2)}\hat{\sigma}_1^-+\hat{\sigma}_2^-\right].
\end{align}
For the braided giant atoms,
\begin{align}
\Delta_{L_1}=&\Delta_{L_2}=\Gamma\sin\left(\phi_1+\phi_2\right),\nonumber\\
\Gamma_1=&\Gamma_2=2\Gamma\left[1+\cos\left(\phi_1+\phi_2\right)\right],\nonumber\\
g_{12}=&\frac{\Gamma}{2}\left[\sin\phi_2+2\sin\phi_1+\sin\left(2\phi_1+\phi_2\right)\right],\nonumber\\
\Gamma_{12}=&\Gamma\left[\cos\phi_2+2\cos\phi_1+\cos\left(2\phi_1+\phi_2\right)\right],\nonumber\\
\Omega_1=&\sqrt{2\Gamma}\alpha\left[1+e^{i(\phi_1+\phi_2)}\right],\nonumber\\
\Omega_2=&\sqrt{2\Gamma}\alpha\left[e^{i\phi_1}+e^{i(2\phi_1+\phi_2)}\right],\nonumber\\
\hat{b}_\text{out}^{(r)}=&\sqrt{\frac{\Gamma}{2}}\left[1+e^{i(\phi_1+\phi_2)}\right]\left(\hat{\sigma}_1^-+e^{i\phi_1}\hat{\sigma}_2^-\right),\nonumber\\
\hat{b}_\text{out}^{(t)}=&\alpha e^{i(2\phi_1+\phi_2)}+\sqrt{\frac{\Gamma}{2}}\left[1+e^{i(\phi_1+\phi_2)}\right]\left(e^{i\phi_1}\hat{\sigma}_1^-+\hat{\sigma}_2^-\right).
\end{align}
For the nested giant atoms
\begin{align}
\Delta_{L_1}=&\Gamma\sin(2\phi_1+\phi_2),\hspace{5pt} \Delta_{L_2}=\Gamma\sin\phi_2,\nonumber\\
\Gamma_1=&2\Gamma\left[1+\cos(2\phi_1+\phi_2)\right], \hspace{5pt} \Gamma_2=2\Gamma\left(1+\cos\phi_2\right),\nonumber\\
g_{12}=&\Gamma\left[\sin\phi_1+\sin\left(\phi_1+\phi_2\right)\right],\nonumber\\
\Gamma_{12}=&2\Gamma\left[\cos\phi_1+\cos\left(\phi_1+\phi_2\right)\right],\nonumber\\
\Omega_1=&\sqrt{2\Gamma}\alpha\left[1+e^{i(2\phi_1+\phi_2)}\right],\nonumber\\
\Omega_2=&\sqrt{2\Gamma}\alpha\left[e^{i\phi_1}+e^{i(\phi_1+\phi_2)}\right],\nonumber\\
\hat{b}_\text{out}^{(r)}=&\sqrt{\frac{\Gamma}{2}}\left\{\left[1+e^{i(2\phi_1+\phi_2)}\right]\hat{\sigma}_1^-+e^{i\phi_1}\left[1+e^{i\phi_2}\right]\hat{\sigma}_2^-\right\},\nonumber\\
\hat{b}_\text{out}^{(t)}=&\alpha e^{i(2\phi_1+\phi_2)}+\hat{b}_\text{out}^{(r)}.
\end{align}

\subsection{Incoherent power spectrum}
We investigate the properties of inelastic scattering using the solution to master equation of the coherently driven system in the weak driving regime. The operator of output field can be decomposed into a sum of a coherent term $\beta^{(m)}$ and an incoherent term $\hat{\zeta}^{(m)}$ as:
\begin{align}
\hat{b}_\text{out}^{(m)}=\beta^{(m)}+\hat{\zeta}^{(m)},
\label{eqn:lin_decom}
\end{align}
where $\beta^{(m)}=\langle\hat{b}_\text{out}^{(m)}\rangle$ ($m=t,r$). The incoherent power spectrum is
\begin{align}
S_\text{incoh}(\omega)=\sum_{m=t,r}\int_{-\infty}^\infty dt e^{-i\omega t}\langle\hat{\zeta}^{(m)\dagger}(t)\hat{\zeta}^{(m)}(0)\rangle_\text{ss},
\end{align}
where $\langle\cdots\rangle_\text{ss}$ denotes the expectation value in the steady state. The total inelastic photon flux is
\begin{align}
F(\omega)=\int_{-\infty}^\infty d\omega S_\text{incoh}(\omega).
\end{align}

\begin{figure}[hbt]
\centering\includegraphics[width=8.5cm,keepaspectratio,clip]{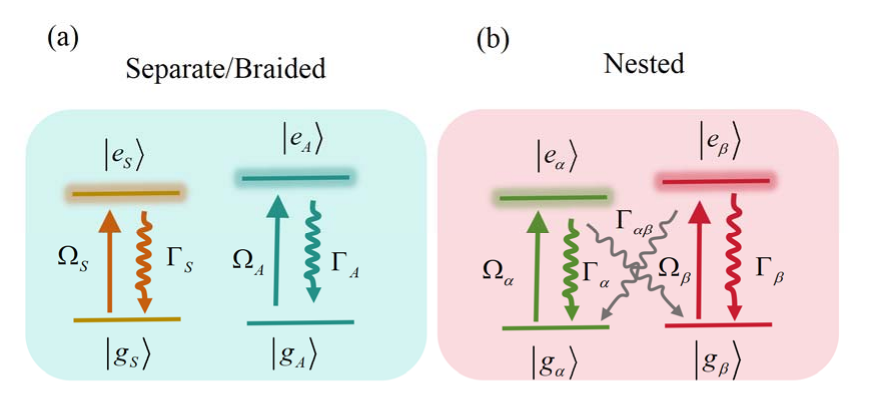}
\caption{Energy diagram illustrating the inelastic power spectra in the three configurations. (a) Independent transitions in the symmetric and antisymmetric basis for the separate and braided configurations. (b) Correlated transitions in the eigen-basis for the nested configuration.}
\label{fig6}
\end{figure}
For the separate and nested giant atoms, it is useful to introduce the symmetric and antisymmetric operators $\hat{\sigma}_{S,A}^-=\left(\hat{\sigma}_1^-\pm\hat{\sigma}_2^-\right)/\sqrt{2}$ due to parity symmetry. The master equation then takes the form
\begin{align}
\frac{d}{dt}\hat{\rho}=-i[\hat{H}_\text{dr},\hat{\rho}]+\sum_{u=S,A}\Gamma_u\mathcal{D}[\hat{\sigma}_u^-]\hat{\rho}
\end{align}
where 
\begin{align}
\hat{H}_\text{dr}=\sum_{u}\Delta_{L_u}\hat{\sigma}_u^+\hat{\sigma}_u^--\frac{i}{2}\sum_u\left[\Omega_u\hat{\sigma}_u^+-\text{H.c.}\right].
\end{align}
Here, $\Delta_{L_{S,A}}=\Delta_{L_1}\pm g_{12}$ are the eigenfrequencies of the symmetry and antisymmetry operators, $\Gamma_{S,A}=\Gamma_1\pm\Gamma_{12}$ are the decay rates, and $\Omega_{S,A}=(\Omega_1\pm\Omega_2)/\sqrt{2}$ are the effective coherent driving strengths. The energy diagram is depicted in Fig.~\ref{fig6}(a). Then, we can derive the following equation
\begin{align}
&\frac{d}{dt}
\left[
\begin{array}{c}
    \langle\hat{\sigma}_u^+\rangle \\
    \langle\hat{\sigma}_u^-\rangle \\
    \langle\hat{\sigma}_u^{ee}\rangle \\
\end{array}
\right]
=\mathbf{M}_u
\left[
\begin{array}{c}
    \langle\hat{\sigma}_u^+\rangle \\
    \langle\hat{\sigma}_u^-\rangle \\
    \langle\hat{\sigma}_u^{ee}\rangle \\
\end{array}
\right]-\frac{1}{2}
\left[
\begin{array}{c}
    \Omega_u^* \\
    \Omega_u \\
    0 \\
\end{array}
\right],\nonumber\\
&\mathbf{M}_u=
\left[
\begin{array}{ccc}
    i\Delta_{L_u}-\Gamma_u/2 & 0 & \Omega_u^* \\
    0 & -i\Delta_{L_u}-\Gamma_u/2 & \Omega_u \\
    -\Omega_u/2 & -\Omega_u^*/2 & -\Gamma_u \\
\end{array}
\right],
\end{align}
where $\hat{\sigma}_u^{ee}=\ket{e_u}\bra{e_u}$. The steady-state solutions for the atomic operators are
\begin{align}
\langle\hat{\sigma}_u^-\rangle=&\frac{\Omega_u}{2}\frac{i\Delta_{L_u}-\Gamma_u/2}{\Delta_{L_u}^2+\Gamma_u^2/4+\abs{\Omega_u}^2/2},\nonumber\\
\langle\hat{\sigma}_u^{ee}\rangle=&\frac{\abs{\Omega_u}^2/4}{\Delta_{L_u}^2+\Gamma_u^2/4+\abs{\Omega_u}^2/2}.
\end{align}
With these steady-state values, we can further calculate the transmission and reflection amplitudes, which are defined as $t=\beta^{(t)}/\alpha$ and $r=\beta^{(r)}/\alpha$. Furthermore, with the use of the quantum regression theory~\cite{scully1999quantum}, it is able to obtain the incoherent power spectra. For the separate giant atoms, the incoherent power spectrum is given by
\begin{align}
S_\text{incoh}(\omega)=&4\Gamma(1+\cos\phi_1)\Big\{\left[1+\cos(\phi_1+\phi_2)\right]\mathcal{S}_S(\omega)\nonumber\\
&+\left[1-\cos(\phi_1+\phi_2)\right]\mathcal{S}_A(\omega)\Big\}.
\end{align}
Similarly, for the braided giant atoms, the incoherent power spectrum is
\begin{align}
S_\text{incoh}(\omega)=&4\Gamma\left[1+\cos(\phi_1+\phi_2)\right]\big\{\left(1+\cos\phi_1\right)\mathcal{S}_S(\omega)\nonumber\\
&+\left(1-\cos\phi_1\right)\mathcal{S}_A(\omega)\big\}.
\end{align}
Here,
\begin{align}
\mathcal{S}_u(\omega)=\text{Re}\left\{
[1,0,0]
(i\omega-\mathbf{M}_u)^{-1}
\left[
\begin{array}{c}
   \langle\hat{\sigma}_u^{ee}\rangle \\
   0 \\
   0 \\
\end{array}
\right]\right\}.
\end{align}
It is obvious that the incoherent power spectra is the sum of two fluorescent light emitted from two effective two-level atoms, as shown in Fig.~\ref{fig6}. The peaks in the spectra correspond to the energy differences of the atoms, while the width is proportional to $\Gamma_u$. The specific values provided in Table~\ref{table1} represent the energy differences and linewidths of these effective two-level atoms.

For the nested giant atoms, the eigen operators can be expressed as
\begin{align}
\hat{\sigma}_\alpha^-=\sin\xi\hat{\sigma}_1^-+\cos\xi\hat{\sigma}_2^-,\hspace{5pt}\hat{\sigma}_\beta^-=-\cos\xi\hat{\sigma}_1^-+\sin\xi\hat{\sigma}_2^-,
\end{align}
where $\xi=\arctan\left[\frac{2g_{12}}{\Delta_{L_2}-\Delta_{L_1}+\sqrt{(\Delta_{L_1}-\Delta_{L_2})^2+4g_{12}^2}}\right]$. The corresponding eigenvalues are
\begin{align}
\Delta_{L_{\alpha/\beta}}=\frac{1}{2}\left[\Delta_{L_1}+\Delta_{L_2}\pm\sqrt{(\Delta_{L_1}-\Delta_{L_2})^2+4g_{12}^2}\right].
\end{align}
The master equation becomes
\begin{align}
\frac{d}{dt}\hat{\rho}=&-i[\hat{H}_\text{dr},\hat{\rho}]+\sum_{v=\alpha,\beta}\Gamma_v\mathcal{D}[\hat{\sigma}_v^-]\hat{\rho}\nonumber\\
&+\Gamma_{\alpha\beta}\sum_{v\neq v^\prime}\left(\hat{\sigma}_v^-\hat{\rho}\hat{\sigma}_{v^\prime}^+-\frac{1}{2}\{\hat{\sigma}_v^+\hat{\sigma}_{v^\prime}^-,\hat{\rho}\}\right),
\end{align}
where 
\begin{align}
\hat{H}_\text{dr}=\sum_{v}\Delta_{L_v}\hat{\sigma}_v^+\hat{\sigma}_v^--\frac{i}{2}\sum_v\left[\Omega_v\hat{\sigma}_v^+-\text{H.c.}\right].
\end{align}
The parameters are defined as
\begin{align}
\Omega_\alpha=&\Omega_1\sin\xi+\Omega_2\cos\xi, \hspace{5pt} \Omega_\beta=-\Omega_1\cos\xi+\Omega_2\sin\xi,\nonumber\\
\Gamma_\alpha=&\Gamma_1\sin^2\xi+\Gamma_2\cos^2\xi+\Gamma_{12}\sin2\xi,\nonumber\\
\Gamma_\beta=&\Gamma_1\cos^2\xi+\Gamma_2\sin^2\xi-\Gamma_{12}\sin2\xi,\nonumber\\
\Gamma_{\alpha\beta}=&-(\Gamma_1-\Gamma_2)\sin\xi\cos\xi-\Gamma_{12}\cos2\xi.
\end{align}
The transitions are illustrated in Fig.~\ref{fig6}(b). Here, $\Gamma_v$ represents the individual relaxation rate, and $\Gamma_{\alpha\beta}$ represents the collective relaxation rate. However, when $\Gamma_{\alpha\beta}\ll\abs{\Delta_{L_\alpha}-\Delta_{L_\beta}}$, the effect of collective relaxation can be ignored. Under this condition, the behavior of the nested giant atoms becomes similar to that of the separate and braided giant atoms. In other words, the peak positions in the incoherent power spectra are determined by the energy differences between eigenvalues, while the widths are proportional to $\Gamma_v$.

\begin{figure}[hbt]
\centering\includegraphics[width=8cm,keepaspectratio,clip]{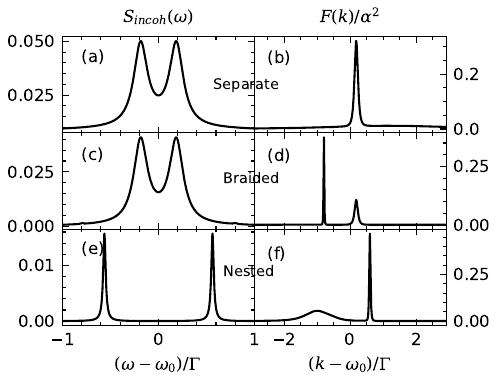}
\caption{The numerical simulation of the incoherent power spectra and total inelastic flux using the master equation approach. Here, $\phi_1=0.25\pi$, $\phi_2=0.85\pi$, $\omega_0=100\Gamma$, and the coherent drive amplitude fulfills $\alpha^2=0.01\Gamma$.}
\label{fig7}
\end{figure}
We numerically plot the incoherent power spectra and the total inelastic photon flux in Fig~\ref{fig7} for the parameters $\phi_1=0.25\pi$, $\phi_2=0.85\pi$. In these plots, the relation of $F/\alpha^2=1-t-r$ is employed, which has been confirmed in Ref.~\cite{PhysRevA.104.063712}. It can observed that the line shape of these figures closely resembles the results shown in Figs.~\ref{fig2} and~\ref{fig3}, confirming the agreement between our analytical results and the numerical calculations. Furthermore, each configuration of giant atoms exhibits distinct spectral features, including resonant peaks and spectral widths. Through analyzing the positions and widths of these spectra, it may be possible to identify the underlying geometry of the giant atom setup.

\subsection{Two-photon differential correlation function}
\begin{figure}[!h]
\centering\includegraphics[width=5cm,keepaspectratio,clip]{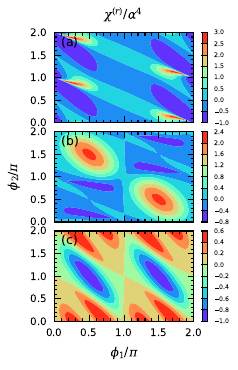}
\caption{Two-photon differential correlation function calculated from the master equation approach. Here, $\omega_0=100\Gamma$ and the coherent drive amplitude fulfills $\alpha^2=0.01\Gamma$.}
\label{fig8}
\end{figure}
Furthermore, the mechanism that leads to bunching and antibunching in the scattered field can be interpreted from the perspective of a superposition of a coherent and fluorescence fields. We follow the numerical procedure similar to the Heitler regime of resonance fluorescence using the master equation approach~\cite{https://doi.org/10.1002/lpor.201900279,LópezCarreño2018,PhysRevLett.108.093602}. Concretely, in Eq.~\eqref{eqn:lin_decom}, the output field consists of both a coherent and incoherent field. To be consistent with the Eq.~\eqref{eqn:chi_L} we consider the zero-delay two-photon differential correlation function, which can be expressed as follows:
\begin{align}
\chi^{(m)}=&\langle\hat{b}_\text{out}^{(m)\dagger2}\hat{b}_\text{out}^{(m)2}\rangle-\langle\hat{b}_\text{out}^{(m)\dagger}\hat{b}_\text{out}^{(m)}\rangle^2\nonumber\\
=&\mathcal{I}_0^{(m)}+\mathcal{I}_1^{(m)}+\mathcal{I}_2^{(m)},
\end{align}
where $\mathcal{I}_0^{(m)}$, $\mathcal{I}_1^{(m)}$, and $\mathcal{I}_2^{(m)}$ represent the powers of $\beta$:
\begin{align}
\mathcal{I}_0^{(m)}=&\langle\hat{\zeta}^{(m)\dagger2}\hat{\zeta}^{(m)2}\rangle-\langle\hat{\zeta}^{m\dagger}\hat{\zeta}^{(m)}\rangle^2,\nonumber\\
\mathcal{I}_1^{(m)}=&4\text{Re}\left(\beta^{(m)*}\langle\hat{\zeta}^{(m)\dagger}\hat{\zeta}^{(m)2}\right),\nonumber\\
\mathcal{I}_2^{(m)}=&2\abs{\beta^{(m)2}}\langle\hat{\zeta}^{(m)\dagger}\hat{\zeta}^{(m)}\rangle+2\text{Re}\left(\beta^{(m)*2}\langle\hat{\zeta}^{(m)2}\rangle\right).
\end{align}

We numerically plot the two-photon differential correlation functions for the reflected fields in Fig.~\ref{fig8}. The shapes of these figures resemble those calculated from the analytical result in Eq.~\eqref{eqn:chi_L}. The subtle differences arise because the numerical calculation using the master equation approach includes more than two-photon scattering processes. The 
agreement between the numerical and analytical results demonstrates the validity of our theoretical framework in capturing the essential features of the system. These correlation functions provide valuable insights into the quantum correlations and interference present in the scattered field.

\bibliography{reference}

\end{document}